\documentclass[11pt]{revtex4}

\usepackage{hyperref}
\usepackage{amsmath,amsfonts,amssymb}
\hypersetup{dvips,dvipdfm,colorlinks=true,urlcolor=magenta,filecolor=magenta,linktoc=page,citecolor=red,linkcolor=blue,bookmarks=true}
\usepackage{graphicx,epstopdf}

\begin{document}

\title{Quantum cosmology with symmetry analysis for quintom dark energy model}

\author{Sourav Dutta$^1$\footnote {sduttaju@gmail.com}}
\author{Muthusamy Lakshmanan $^2$\footnote {lakshman.cnld@gmail.com}}
\author{Subenoy Chakraborty$^3$\footnote {schakraborty.math@gmail.com (Corresponding Author)}}
\affiliation{$^1$Department of Mathematics, Dr Meghnad Saha College, Itahar, Uttar Dinajpur-733128, West Bengal, India\\$^2$ Centre for Nonlinear Dynamics, Bharathidasan University, Tiruchirapalli - 620 024, India\\$^3$Department of Mathematics, Jadavpur University, Kolkata-700032, West Bengal, India}


\begin{abstract}
Quantum cosmology with quintom dark energy model has been investigated in the present work using symmetry analysis of the underlying physical system. In the background of the flat FLRW model quintom cosmological model has been studied using Noether symmetry and appropriate conserved charge is obtained. The Wheeler-DeWitt (WD) equation is constructed on the minisuperspace and solutions are obtained using conserved charge.
	
\end{abstract}

\maketitle

\section{Introduction}
A series of observational evidences \cite{r1,r2,r3,r4,r5} since the end of the last century put new windows to cosmology. To accommodate the present accelerated expansion of the Universe in the framework of standard cosmology, cosmologists are pursuing two paths:1) Einstein gravity is replaced by some modified gravity theory or 2) some exotic matter (known as dark energy) is added to the normal matter. This mysterious matter component (i.e., dark energy (DE)) is completely unknown so far except for its large negative pressure violating (at least) strong energy condition. Initially, cosmologists opted for cosmological constant--the simplest candidate for DE. But due to two well known drawbacks, namely the cosmological constant problem and coincidence problem \cite{r6.1}, the dynamical DE models \cite{r6.2,r6.3,r6.4,r6.5} are widely used. Most of the dynamical DE models can be classified into two groups--one having equation of state parameter grater than $-1$ i.e., in quintessence era while others in phantom domain with equation of state parameter less than $-1$. But it is hard to find one which may cross the phantom barrier. In this context  k-essence DE model \cite{r6.6} or a (phantom) scalar tensor theory with exponential potential \cite{r6.61} may have equation of state parameter in both the eras but it is hard to cross the cosmological constant during evolution.\\
On the other hand Planck's data since 2013 \cite{r6.7,r6.8}indicate that the equation of state parameter for DE should be very close to phantom  barrier and it may be in any one of these two regions within the error bar. In particular Planck' 2018 \cite{r6.9}data shows $w=-1.028 \pm 0.031$ at $68\%$ confidence limit.\\
Further, from a theoretical point of view there exists a no--go theorem \cite{r6} which does not allow a single scalar field model to cross over the phantom barrier. Thus based on the recent observational evidences and theoretical limitation, multiscalar field models appear to be very useful as DE candidate. Quintom DE model \cite{r6.10,r6.11,r6.12,r6.13,r6.14,r6.15} is such a simple multiscalar field model. Here two scalar fields, of which one is normal and the other is ghost in nature, are minimally coupled to gravity and have a coupled self interacting potential.\\

From a different perspective, geometric symmetries related to space-time and symmetries of the physical system are very much important to analyze any physical motion. In fact, the conserved quantities (known as conserved charges) in the context of Noether symmetry can be used as a selection criteria to distinguish similar physical processes (for example different DE models). Further, mathematically the first integral (i.e., the Noether integral) in Noether symmetry approach has been used as a tool for simplification of a system of differential equations or for studying the integrability of the system. Also it is possible to check the self-consistency of phenomenological physical models using Noether symmetries. Moreover, one can constrain the physical parameters involved in a physical system by symmetry analysis \cite{r7}. As an example, it is worthy to mention that in the recent past the self interacting coupled potential of the quintom model has been evaluated \cite{rs} (instead of choosing phenomenologically) by imposing Noether symmetry to the physical model. In recent years considerable amount of work \cite{r8, r9, r10, r11, r12, r13, r14} has been done with symmetry analysis for models in Riemannian spaces.\\

Symmetry analysis has been extensively used in the context of quantum cosmology. The WD equation is constructed (with suitable operator ordering) on the minisuperspace and it is related to the existence of Lie point symmetries. Also the Noether symmetries are very much important in quantum cosmology as they provide a subset of the general solutions of the WD equation having oscillatory behaviours with crucial physical meaning \cite{r15, r16}. Further, Hartle criterion   can be related to the Noether symmetries and in minisuperspace they identify those classical trajectories  which are solutions of the cosmological evolution equations \cite{r17, r18}. In otherwords, Noether symmetries provide a bridge between quantum cosmology and classically observable Universes.\\

In the present work symmetry analysis has been extensively used to formulate quantum cosmology for the quintom DE model. By using symmetry analysis the WD equation is constructed over the minisuperspace. The symmetry analysis of the WD equation is shown
to be related to the conformal Killing vector field or homothetic vector field of the minisuperspace. The conserved charge associated with Noether symmetry of the WD equation has shown to have an active role in solving WD equation. The paper is organized as follows: A general prescription for conformal and Noether symmetries is presented in Section-II, whereas Section-III describes the formation of the Wheeler-DeWitt equation and Minisuperspace. Section-IV deals with the results of WD equation using Noether symmetry and finally the paper ends with a short discussion in Section-V. 
\section{Conformal symmetry and Noether symmetry: A general prescription}
Conformal invariance is very much important for quantum cosmology due to its rich geometrical structure. If a vector field $X^a$ in space with metric $g_{ij}$ satisfies the relation
\begin{equation}
\mathcal{L}_{_{\overrightarrow{X}}} g_{ij}=\mu (x^k) g_{ij}, \label{2.1}
\end{equation}
where $\mu(x^k)$ is an arbitrary function of $x^k$, then $\overrightarrow{X}$ is called a conformal killing vector of the space. Depending on the choice of the arbitrary function $\mu (x^k)$ the vector field $X^a$ has different names:
\begin{eqnarray}
~~~~~~~\mbox{a)}~~~ \mu (x^k)&=& \mu_0 (\neq 0),  \mbox{a constant, then}\nonumber\\
~~~\mathcal{L}_{_{\overrightarrow{X}}} g_{ij} &\propto& g_{ij} : \overrightarrow{X} \mbox{ is a homothetic vector field.}\nonumber
\end{eqnarray}

\begin{eqnarray}
\mbox{b)}~~~ \mu (x^k)&=& 0\mbox{, then}\nonumber\\
~~~\mathcal{L}_{_{\overrightarrow{X}}} g_{ij} &=& 0 : \overrightarrow{X} \mbox{ is a Killing vector field.}\nonumber
\end{eqnarray}
In a space if $\exists$ two metrics $g$ and $\tilde{g}$ such that
\begin{equation}
\tilde{g}_{ij}=\xi^2 (x^k) g_{ij}, \label{2.2}
\end{equation}
for some scalar function $\xi$, then the above two metrics are said to be conformal.

The set of all conformal killing vectors form an algebra, known as conformal algebra. So two conformally related metrics have same conformal algebra (CA)
and the conformal factors are related by the relation :
\begin{equation}
\bar{\mu}(x^k)=\mu (x^k)+\mathcal{L}_{_{\overrightarrow{X}}} (\ln \xi), \label{2.3}.
\end{equation}
Similarly, the set of all homothetic vector fields and the set of all killing vector fields form algebras and are respectively termed as homothetic algebra (HA) and Killing algebra (KA) respectively. Also these two algebras are subalgebras of the conformal algebra, i.e.,
\begin{equation}
KA\subseteq HA \subseteq CA. \label{2.4}.
\end{equation}
In an $n(>2)$ dimensional space of constant curvature the dimension of these three algebras are respectively $\frac{n(n+1)}{2},~\frac{n(n+1)}{2}+1,~\frac{(n+1)(n+2)}{2}.$ However due to the relation (\ref{2.3}), the subalgebras of two conformally related metrics are not identical.

From a physical point of view, it will be interesting to study  systems having Lagrangians which are related conformally. It has been found \cite{r19} that  the equations of motion (i.e., Euler-Lagrange equations) of two conformally related Lagrangians transform covariantly under a conformal transformation provided the total energy (i.e., Hamiltonian) is zero. So equivalently any two physical systems having vanishing energy are conformally related and the corresponding evolution equations are conformally invariant.

This idea of conformally equivalent Lagrangians has been subsequently extended to scalar field cosmology in general Riemannian space \cite{r19}. However, in quantum cosmology the total energy of the system is zero due to Hamiltonian constraint, and so in the present context it is useful to study conformally invariant systems with respect to equations of motion. But from the point of view of Noether symmetries two conformally related physical systems are not identical as the homothetic algebras (related to the Noether symmetries) for two conformally related metrics are distinct.

The Noether symmetry approach is very much relevant to identify conserved quantities of a physical system. Here the symmetry vector is defined on the tangent space of configurations: $TQ \equiv \{q, \dot{q}\}$ as
\begin{equation}
\overrightarrow{X}=\alpha(q)\frac{\partial}{\partial q}+\dot{\alpha}(q)\frac{\partial}{\partial \dot{q}}. \label{2.5}
\end{equation}
Then according to Noether's theorem \cite{n6, n7}
\begin{equation}
\mathcal{L}_{_{\overrightarrow{X}}} L=\overrightarrow{X}L=\alpha(q)\frac{\partial L}{\partial q}+\dot{\alpha}(q)\frac{\partial L}{\partial \dot{q}}=0. \label{2.6}
\end{equation}
Here the Lagrangian $L$ is defined over $TQ$. The above Noether condition (\ref{2.6}) implies a constant of motion for the Lagrangian or equivalently the phase flux is conserved along the vector field $\overrightarrow{X}$. From the point of view of quantum cosmology the above Noether condition in Hamiltonian formulation takes the form 
\begin{equation}
\mathcal{L}_{_{\overrightarrow{u}}} \mathcal{H}=0, \label{2.7}
\end{equation}
with $\overrightarrow{u}=\dot{q}\frac{\partial}{\partial q}+\ddot{q}\frac{\partial}{\partial \dot{q}}$

So in minisuperspace models the above symmetry condition gives appropriate interpretation of the wave function of the Universe. Due to the above Noether symmetry the conserved canonically conjugate momenta are
\begin{equation}
\Pi_i \equiv \frac{\partial L}{\partial q^i}=i_{_{\overrightarrow{x_i}}} \theta_L=\Sigma_i~,~~ i=1, 2,.......m \label{2.8}
\end{equation}
where the positive integer `$m$' denotes the number of symmetries. Writing the operator version of the momentum variable one has the quantum version of the above symmetry condition as  
\begin{equation}
-i \partial_{q^l}|\psi>=\Sigma_l|\psi>, l = 1, 2, ---,m\label{2.9}
\end{equation}
Geometrically, the corresponding symmetry condition implies a translation along the $q^l$-axis. Thus for real conserved quantities (i.e., $\Sigma_{l}$ to be real) the wave function has the oscillatory behaviour along the direction of symmetries as 
\begin{equation}
|\psi>=\sum_{l=1}^{m} e^{i \sum _l q^l}\bigg|\phi(q^k)\bigg>,~k<n \label{2.10}
\end{equation}
where `$n$' is the dimension of the minisuperspace of which `$k < n$' directions have no symmetry and $\big|\phi(q^k)\big>$ is the non-oscillatory part of the wave function associated with the subspace of the minisuperspace having no symmetry. Thus due to Noether symmetry, the corresponding reduction procedure gives oscillatory behaviour of the solution of the WD equation along the direction of symmetries. Further, due to Hartle \cite{rh} the wave function of the Universe with conserved momenta due to symmetry, corresponds to trajectories as classical cosmological solutions. Therefore, it is reasonable 
to think that the oscillatory part of the solution of the WD equation corresponds to conserved momenta associated with Noether symmetries along those directions.

\section{Quantum cosmology in quintom dark energy model: Formation of Wheeler-DeWitt equation and minisuperspace}
The Lagrangian of the quintom dark energy model having scalar fields $\phi$ and $\sigma$ in the background of flat FLRW model is given by \cite{rs}
 \begin{equation}
 L\big(a, \dot{a}, \phi, \dot{\phi}, \sigma, \dot{\sigma}\big)=-3a \dot{a}^2+a^3\bigg(\frac{1}{2} \dot{\phi}^2-\frac{1}{2} \dot{\sigma}^2-V(\phi, \sigma)\bigg), \label{3.1}
 \end{equation}
where $V(\phi, \sigma)$ is the coupled self interacting potential.\\

The Friedmann equations are
\begin{equation}
3\frac{\dot{a}^2}{a^2}=\frac{1}{2} \dot{\phi}^2-\frac{1}{2} \dot{\sigma}^2+V(\phi, \sigma),\label{3.2}
\end{equation}
and
\begin{equation}
2\frac{\ddot{a}}{a}+\frac{\dot{a}^2}{a^2}=-\frac{1}{2} \dot{\phi}^2+\frac{1}{2} \dot{\sigma}^2+V(\phi, \sigma).\label{3.3}
\end{equation}
The matter conservation equations are
\begin{equation}
\ddot{\phi}+3H\dot{\phi}+\frac{\partial V}{\partial \phi}=0,\label{3.4}
\end{equation}
and
\begin{equation}
\ddot{\sigma}+3H\dot{\sigma}-\frac{\partial V}{\partial \sigma}=0,\label{3.5}
\end{equation}
where $H$ is the usual Hubble parameter defined by $H \equiv \frac{\dot{a}}{a}$.\\
The three dimensional configuration space is described by the co-ordinates $\{a, \phi, \sigma\}$. So the momenta conjugate to these coordinates are 
\begin{equation}
p_a=\frac{\partial L}{\partial \dot{a}}=-6a\dot{a},~p_{\phi}=\frac{\partial L}{\partial \dot{\phi}}=a^3 \dot{\phi},~p_{\sigma}=\frac{\partial L}{\partial \dot{\sigma}}=-a^3 \dot{\sigma}.\label{3.6}
\end{equation}
So the above Hamiltonian constraint in terms of the momenta takes the form
\begin{equation}
\mathcal{H}=-\frac{1}{12} \frac{p_a^2}{a}+\frac{1}{2a^3} p_{\phi}^2-\frac{1}{2a^3}p_{\sigma}^2+a^3V(\phi, \sigma)=0.\label{3.7}
\end{equation}
The Hamilton's equations of motion are
\begin{eqnarray}
\dot{a}&=&-\frac{\partial \mathcal{H}}{\partial p_a}=\frac{1}{6}\frac{p_a}{a}~,\dot{\phi}=-\frac{\partial \mathcal{H}}{\partial p_{\phi}}=-\frac{p_{\phi}}{a^3}~,\dot{\sigma}=-\frac{\partial \mathcal{H}}{\partial p_{\sigma}}=\frac{p_{\sigma}}{a^3},\nonumber\\
\dot{p_a}&=&\frac{\partial \mathcal{H}}{\partial a}=\frac{1}{24}\frac{p_a^2}{a^2}-\frac{3}{2a^4}p_{\phi}^2+\frac{3}{2a^4}p_{\sigma}^2+3a^2V(\phi, \sigma)~,\dot{p_{\phi}}=\frac{\partial \mathcal{H}}{\partial \phi}=a^3\frac{\partial V}{\partial  \phi}~,\dot{p_{\sigma}}=\frac{\partial \mathcal{H}}{\partial \sigma}=a^3\frac{\partial V}{\partial \sigma}\label{3.8}
\end{eqnarray}
The above Lagrangian of the quintom model can be interpreted as follows: The first three terms in the Lagrangian are identified with the kinetic part while the dynamic part is characterized by the potential term. Moreover, the kinetic part may be geometrically described as a $3D$ pseudo Riemannian space having line element
\begin{equation}
dS_3^2=-6a(da)^2+a^3 d\phi^2-a^3 d\sigma^2.\label{3.9}
\end{equation}
This $3D$ Lorentzian manifold can be identified as the configuration space and is also termed as minisuperspace in quantum cosmology.

The wave function of the Universe in quantum cosmology is the solution of the WD equation which is essentially the operator version of the Hamiltonian constraint. In general, the WD equation is a second order partial differential equation of hyperbolic type. Also it is the Klein-Gordon (KG) equation corresponding to conformal Laplacian operator over the minisuperspace. In the present $3D$ minisuperspace model the explicit form of the WD equation is
\begin{equation}
	\bigtriangleup \psi+\frac{R}{8}+V. \psi=0,\label{3.10}
\end{equation}
where 
\begin{eqnarray}
\bigtriangleup&=& \frac{1}{\sqrt{|g|}}\frac{\partial}{\partial x^i}\bigg(\sqrt{|g|}\frac{\partial}{\partial x^i}\bigg) \nonumber\\
&\equiv& -\frac{1}{a^3}\bigg(\frac{\partial^2}{\partial \phi^2}+\frac{\partial}{\partial \phi}\bigg)+\frac{1}{6a}\frac{\partial^2}{\partial a^2}+\frac{1}{a^3}\frac{\partial^2}{\partial \sigma^2},\nonumber
\end{eqnarray}
is the conformal Laplacian operator, $R$ is the Ricci scalar and $g$ is the metric tensor of the minisuperspace.

As the Lie point symmetries of the KG equation (\ref{3.10}) are connected to the conformal algebra of the minisuperspace metric $g_{ij}$ so the Lie point symmetry vector can be expressed as \cite{r15}
 \begin{equation}
 \overrightarrow{X}=\xi^i(x^k)\partial_i+\bigg[-\frac{1}{2} \lambda \psi+a_0 \psi\bigg]\partial \psi,\label{3.11}
 \end{equation}
where $a_0$ is a constant, $\xi^i$ is the conformal killing vector and $\lambda(x^k)$ is the conformal factor of the present $3D$ minisuperspace. As a consequence, the potential function is determined by the Lie point symmetry as
 \begin{equation}
 \mathcal{L}_{_{\overrightarrow{\xi}}}V+2\lambda V=0.\label{3.12}
 \end{equation}
In WKB approximation, one writes the wave function of the Universe as $\psi(x^k) \sim e^{is(x^k)}$ so that the above WD equation (\ref{3.10}) reduces to the (null) Hamilton-Jacobi equation as
 \begin{equation}
-\frac{1}{12a}\bigg(\frac{\partial S}{\partial a}\bigg)^2+ \frac{1}{2a^3}\bigg(\frac{\partial S}{\partial \phi}\bigg)^2-\frac{1}{2a^3}\bigg(\frac{\partial S}{\partial \sigma}\bigg)^2+a^3 V(\phi, \sigma)=0.\label{3.13}
 \end{equation}
From the point of view of Liouville integrability there should be at least 2 independent Lie point symmetries to the WD equation of the present $3D$ WD equation. Hence the solution of the WD equation can be expressed  in terms of zero order invariants of these Lie point symmetries as 
 \begin{equation}
\psi(\tilde{x}^3, \hat{x}^j)=\phi(\tilde{x}^3) \exp\bigg[\sum_{j=1}^2 \int \bigg\{-\frac{\lambda}{2}-Q_j\bigg\}d\tilde{x}^j\bigg].\label{3.14}
 \end{equation}
Here the constants of motion $Q_j$, are along the symmetry directions and $\phi(\tilde{x}^3)$ satisfies a linear second order ordinary differential equation. Hence the Lie point symmetries of the WD equation which are determined from the conformal Killing vector of the minisuperspace are useful to reduce or solve the WD equation. Due to distinct conformal Killing algebra of two conformally related matrices the Lie point symmetry vectors (corresponding to conformal algebra) will not be identical and hence the solutions will be different.

\section{Noether symmetry and quintom model} 
In the present quintom cosmological model the configuration space (i.e., minisuperspace) is three dimensional $\{a, \phi, \sigma\}$. So the Noether symmetry vector 
\begin{equation}
\overrightarrow{X}=\mu\frac{\partial}{\partial a}+\lambda\frac{\partial}{\partial \phi}+\xi\frac{\partial}{\partial \sigma}+\dot{\mu}\frac{\partial}{\partial \dot{a}}+\dot{\lambda}\frac{\partial}{\partial \dot{\phi}}+\dot{\xi}\frac{\partial}{\partial \dot{\sigma}},\label{4.6}
\end{equation}
acts on the tangent space. As usual $\mu=\mu(a, \phi, \sigma ),~~\lambda=\lambda(a, \phi, \sigma),~~\xi=\xi(a, \phi, \sigma)$
and 
$\dot{\mu}=\frac{\partial \mu}{\partial a}\dot{a}+\frac{\partial \mu}{\partial \phi}\dot{\phi}+\frac{\partial \mu}{\partial \sigma}\dot{\sigma}$ and so on.

A detailed study of the Lie and Noether symmetries associated with the coupled equations (\ref{3.2}--\ref{3.5}) for the present quintom dark energy model is done in ref.\cite{rs}. The coefficients of the above Noether symmetry vector (\ref{4.6}) are determined from the  constituent equations of the Noether symmetry condition (\ref{2.6}) as \cite{rs}
\begin{equation}
\mu=0,~\lambda=k\sigma+d_1,~\xi=k\phi+d_2,\label{4.7}
\end{equation}
Also the constituent equations determine the coupled potential of the quintom scalar fields in the forms \cite{rs} 
\begin{equation}
V(\phi, \sigma)=v_1 \bigg[\frac{k}{2}\big(\phi^2-\sigma^2 \big)+d_2 \phi-d_1 \sigma \bigg],\label{4.8}
\end{equation}
where $k, d_1$ and $d_2$ are constants of integration.

The above potential has the following features:

(a)~If $k=0$ and $d_2=d_1$, then the potential is antisymmetric in nature under interchange of $\phi$ and $\sigma$ (as well as reflection symmetry) and it is invariant under the translation $\phi \rightarrow \phi+l, \sigma \rightarrow \sigma+l$ ($l$ is arbitrary constant).

(b)~ If $k\neq 0$ and $d_1=d_2=0$, then also the potential is antisymmetric under exchange of $\phi$ and $\sigma$, however it has reflection symmetry too (i.e, $\phi \rightarrow -\phi, \sigma \rightarrow -\sigma$).

(c)~ $k=0,~d_1=-d_2\neq 0$, then the potential is a symmetric function under exchange of $\phi$ and $\sigma$. \\

Usually, it is well known in this kind of problem to fix the appropriate scalar potential with single or multi--fields is that does not exist an underline principle which will specify uniquely the potential. So people employ some particular characteristics for obtain this, for example solving the Klein-Gordon equation in quadrature form \cite{a}, algebraic methods \cite{b}, employing Bohm's idea \cite{c}, Lie symmetry invariance \cite{r8} (see also \cite{d, e, f}). In \cite{r8} the authors have applied the method \cite{g} of introducing a flat metric and a potential function by decomposing the Lagrangian into kinetic energy part (termed as kinetic-energy metric) and the potential energy part (which defines the potential) while in the present work the coupled potential of the scalar fields is determined (in ref \cite{rs}) purely from the Noether symmetry condition.\\ 

In general, the conserved quantity associated with Noether's symmetry is the energy-momentum tensor. However, for space-time having time-like Killing vector field one can have conserved energy. So usually FLRW space-time model does not have conserved energy. But for the present model as the Lagrangian does not depend explicitly on time and so in analogy with point-like Lagrangian it is possible to have a conserved energy (i.e., the Hamiltonian). Thus one has two conserved quantities, namely the conserved charge

\begin{equation}
Q=a^3 k\big(\sigma \dot{\phi}-\dot{\sigma} \phi \big)+a^3 \big( d_1 \dot{\phi}-d_2 \dot{\sigma}\big),\label{4.9}
\end{equation}
and conserved energy
\begin{equation}
E=-3a \dot{a}^2+\frac{1}{2}a^3 \dot{\phi}^2-\frac{1}{2} a^3\dot{\sigma}^2+a^3 V(\phi, \sigma).\label{4.10}
\end{equation}
(There is a conserved current associated with Noether symmetry and its time component when integrated over the spatial volume gives the Noether charge. But the Lagrangian in the present case depends only on time and so one gets the Noether charge).

Further, to simplify the evolution equations a transformation of the augmented space variables $(a, \phi, \sigma)\rightarrow (u, v, w),$ is determined in ref. \cite{rs} as

{\bf I: $k\neq 0$}

\begin{eqnarray}
e^{ku}&=& k\big(\phi+\sigma \big)d_1+d_2,
\nonumber
\\
v&=& a,
\nonumber
\\
w&=& \frac{k}{2}\big(\phi^2-\sigma^2 \big)+d_1 \phi-d_2 \sigma.\label{4.11}
\end{eqnarray}

{\bf II: $k= 0$}

\begin{eqnarray}
u&=& \frac{\phi+\sigma}{d_1+d_2},
\nonumber
\\
v&=& a,
\nonumber
\\
w&=& d_1 \phi-d_2 \sigma.\label{4.12}
\end{eqnarray}
so that one of the variables $(u)$ becomes cyclic and the Lagrangian takes the form
\begin{equation}
L_T=\left\{
\begin{array}{lll}
-3v\dot{v}^2+v^3\big(\dot{w} \dot{u}-w k \dot{u}^2-v_1 w\big), & \mbox{ for } k\neq 0, d_1=d_2=0\\
-3v\dot{v}^2+v^3\big(\dot{w} \dot{u}-v_1 w\big), & \mbox{ for } k=0, d_1=d_2=d\\
-3v\dot{v}^2+v^3\big(\dot{w} \dot{u}-v_1 w\big) & \mbox{ for } k= 0, d_2=-d_1=d\\
\end{array}
\right.,\label{4.13}
\end{equation}
The conjugate momenta corresponding to the new variables $(u, v, w)$ are 
\begin{eqnarray}
\Pi_u&=& -2kw v^3 \dot{u}+v^3 \dot{w},~~k \neq 0\nonumber\\
&=&v^2 \dot{w},~k=0,\nonumber\\
\Pi_v&=& -6v \dot{v}\nonumber\\
\Pi_w &=&v^3 \dot{u}\nonumber.
\end{eqnarray}
Hence the Hamiltonian in the new variables takes the form
\begin{equation}
\mathcal{H}=\left\{
\begin{array}{lll}
-\frac{1}{12}\frac{\Pi_v^2}{v}+\frac{\Pi_u \Pi_w}{v^3}+\frac{kw}{v^3}\Pi_w^2+v_1 v^3 w, & \mbox{ for } k\neq 0,\\
-\frac{1}{12}\frac{\Pi_v^2}{v}+\frac{\Pi_u \Pi_w}{v^3}+v_1v^3w, & \mbox{ for } k=0.\\
\end{array}
\right..\label{4.14}
\end{equation}
Due to the cyclic nature of $u$, we have
\begin{equation}
\Pi_u=v^3 \dot{w}-2kwv^3 \dot{u}=Q, ~\mbox{a constant}\label{4.15}
\end{equation}
We shall now focus our attention to the quantization of the model. In this context the fundamental equation is the WD equation $\hat{\mathcal{H}}\psi(u, v, w)=0$, where $\hat{\mathcal{H}}$ is the operator version of the Hamiltonian (\ref{4.14}) and $\psi(u, v, w)$ is known as the wave function of the Universe. An important issue in the quantization scheme is the factor-ordering problem, related to the ordering of a variable and its conjugate momentum. In the present Hamiltonian (\ref{4.14}) there is a product of $v$ and $\Pi_v$ in the first term (both for $k=0$ and $k\neq 0$) and $w$ and $\Pi_w$ (for $k\neq 0$) in the third term. So the ordering consideration should be taken into account. Thus with the usual operator conversion  $\Pi_u \rightarrow -i \partial_u,~\Pi_v \rightarrow -i \partial_v,~\Pi_w \rightarrow -i \partial_w$, one has the four parameter family of Wheeler DeWitt (WD) equation
\begin{equation}
\bigg[-\frac{1}{12}\frac{1}{v^{l_1}}\frac{\partial}{\partial v}.\frac{1}{v^{m_1}}\frac{\partial}{\partial v}\frac{1}{v^{n_1}}-\frac{1}{v^3}\frac{\partial^2}{\partial u \partial w}-\frac{k}{v^2}w^{l_2}\frac{\partial}{\partial w} w^{m_2}\frac{\partial}{\partial w}w^{n_2}-v_1v^3w\bigg]\psi(u, v, w)=0,\label{4.15.1}
\end{equation}
with real number triplets $(l_1, m_1, n_1)$ and $(l_2, m_2, n_2)$ satisfying $l_1+m_1+n_1=1=l_2+m_2+n_2$. As there are infinite possible choices of the above real triplets so there are infinite possibilities of ordering. Further the factor ordering problem can not be resolved using semi-classical limit as the Hamilton-Jacobi equation so obtained (Using $\psi=\exp (is))$ does not  regard to the above triplets. However, there are some preferred choices for the triplets namely (i) $l_1=2, m_1=-1, n_1=0$, $l_2=2, m_2=-1, n_2=0$ which gives Laplace Beltrami operator and is known as D'Alembert operator ordering, (ii) $l_1=n_1=0, m_1=1, l_2=n_2=0, m_2=1$ is known as Vilenkin ordering, (iii) $l_1=1, m_1=n_1=0, l_2=1, m_2=n_2=0$, no ordering \cite{40}. Note that though factor ordering affects the behaviour of the wave function, yet semi-classical calculations will be unaffected due to factor ordering \cite{41}. In the present case, we shall for simplicity choose the third option (i.e., no ordering) so that the WD equation takes the form  
\begin{equation}
\bigg[-\frac{1}{12}\frac{\partial^2}{\partial v^2}+\frac{1}{v^3} \frac{\partial^2}{\partial w \partial u}+\frac{kw}{v^3}\frac{\partial^2}{\partial w^2}+v_1 v^3 w\bigg]\psi(u, v, w)=0.\label{4.16}
\end{equation}
Usually, the wave function of the Universe is the general solution of the above WD equation which is constructed from the superposition of its eigenfunctions as
\begin{equation}
\Psi(u, v, w)=\int \int dQ ~d\lambda~ A(Q)~C(\lambda)\psi(Q, \lambda, u, v, w), \label{4.16.1}
\end{equation}
where $\psi$ is a eigenfunction of the WD  equation and $A(Q)$, $C(\lambda)$ are the weight functions. Note that by suitable choices of the weight functions one may form desire wave packet. In quantum cosmology it is desirable to construct wave function so that they agree with classical model i.e., one has to construct a coherent wave packet with good asymptotic behaviour in the minisuperspace, packing around the classical trajectory.\\

Further it is interesting to examine whether the evolution of the dynamical variables can be predicted through the wave function in the frame work of the present cosmological model. Usually, a consistent quantum cosmological model should predict the classical solution at late time but have a distinct (different from classical solution) solution in early era which is singularity free. \\
Now the operator version of the conserved equation (\ref{4.15}), i.e.,
$$-i\frac{\partial \psi}{\partial u}=Q\psi.$$
With the separation of $\psi$ as $\psi(u, v, w)=A(u) \phi(v,w)$ gives $A(u)=\exp (i \Sigma_0 u)$, so that equation (\ref{4.16}) becomes
\begin{equation}
\bigg[-\frac{1}{12}\frac{\partial^2}{\partial v^2}+\frac{i}{v^3} \frac{\partial}{\partial w}+\frac{kw}{v^3}\frac{\partial^2}{\partial w^2}+v_1 v^3 w\bigg]\phi(v, w)=0.\label{4.17}
\end{equation}
It is hard to solve  this inhomogeneous linear partial differential equation in two variables $v$ and $w$. However, by using a typical approach one may have the following particular solutions:\\
 
{\bf (a)$k=0$:}
\begin{equation}
\phi(v, w)=\phi_0 e^{-\frac{i\lambda w}{Q}}.v^{\frac{1}{2}}.\chi(v).\label{4.20}
\end{equation}
~~~{\bf (b)$k\neq0$:}
\begin{equation}
\phi(v, w)=\bigg\{\phi_1 w^{\frac{\big(\frac{iQ}{k}+1\big)}{2}}J_{\alpha}\bigg[2\sqrt{\frac{\lambda}{k}w}\bigg]+\phi_2 w^{-\frac{\big(\frac{iQ}{k}+1\big)}{2}}J_{-\alpha}\bigg[2\sqrt{\frac{\lambda}{k}w}\bigg]\bigg\}v^{\frac{1}{2}}\chi(v),\label{4.21}
\end{equation}
with \begin{eqnarray}
\chi(v)&=& c_1 v^{m_1}+c_2 v^{m_2},~~\mbox{if} ~\lambda < \frac{1}{48},\nonumber\\
&=& c_1 \cos (m_3 \ln v)+c_2 \sin (m_3 \ln v),~~\mbox{if}~ \lambda > \frac{1}{48},\nonumber\\
&=&c_1+c_2 \ln v,~~\mbox{if}~ \lambda = \frac{1}{48},\label{4.22}
\end{eqnarray}
where $\phi_0, \phi_1, \phi_2, c_1$ and $c_2$ are arbitrary  integration constants and
\begin{equation}
m_1, m_2=\frac{1\pm \sqrt{1-48\lambda}}{2},~m_3=\frac{\sqrt{48\lambda-1}}{2},
\end{equation}
and $J_{\alpha}$ is the Bessel function of complex order $\alpha (=1-\frac{iQ}{k})$ .\\
The above solution for the wave function  shows purely oscillatory nature (in all the variables) for the choice $k=0, \lambda=\frac{1}{48}$, but it can not hold near the big bang singularity. On the other hand, the solution for $\lambda<\frac{1}{48}$ (for both $k=0$ and $k \neq 0$) is well defined near the big bang singularity so that it is possible to have the wave function of the Universe in the quantum era near the big bang singularity. Further, a good characteristic for appropriate wave function have a damping behaviour for large scale factor. In the present solution (\ref{4.22}) with $\lambda>\frac{1}{48}$ have the above characteristic.
\section{summary}
The present work deals with quantum cosmology for quintom dark energy model from the point of view of Lie and Noether symmetries. In the background of homogeneous and isotropic FLRW space-time, the full quantum theory is described on the infinite dimensional superspace. However for practical purpose we shall restrict to minisuperspace which for the present physical model is a three dimensional Lorentzian manifold. The evolution of the wave function is described by the WD equation, a hyperbolic partial
differential equation over the minisuperspace. The conformal algebra of the Lorentzian metric of the minisuperspace is shown to be related to the Lie point symmetries of this WD equation. The conformal Killing vector field of this minisuperspace is also related to the symmetry vector of the Lie symmetry. On the other hand, the Noether symmetry of the WD equation is related to the homothetic vector field of the minisuperspace. However, due to conformal transformation one has identical conformal algebra with distinct Killing and homothetic subalgebras. As a result the quantum cosmology so constructed by conformal transformations will be distinct. Further it is worthy to mention that Noether symmetry is related to the homothetic vector field of the minisuperspace. \\

Moreover, Noether symmetry analysis to the minisuperspace helps to a great extent in solving the WD equation. The quantum version of the conserved momentum indicates oscillatory solution of the WD equation and it gives the semi-classical limit of quantum cosmology. Subsequently, the non-oscillatory part of the WD equation is solved with a typical technique. The variable `$v$' is nothing but the scale factor `$a$' and so the solution $\chi(v)$ in equation (\ref{4.22}) can be extended to $a\rightarrow 0$ only for the first choice with $\lambda<\frac{1}{48}$, otherwise $\chi(v)$ can not be defined very close to big-bang singularity. Hence from quantum cosmological point of view the first choice for $\chi(v)$ with $\lambda<\frac{1}{48}$ is physically realistic and it can described the quantum era of evolution at the early phase avoiding the big-bang singularity. Therefore, it is reasonable to conclude that symmetry analysis of the minisuperspace mostly characterizes quantum cosmology of a physical system and helps to a great extent in solving the WD equation.

\section*{Acknowledgments}
 Author ML thanks the
Department of Science and Technology, Government of India for the award of a DST-SERB Distinguished Fellowship. S.C. thanks Science and Engineering Research Board (SERB) for awarding
MATRICS Research Grant support (File No. MTR/2017/000407) and RUSA 2.0 of Jadavpur University.

\end{document}